# Complete amplitude and phase control of light using broadband holographic metasurface

*Gun-Yeal Lee[1], Gwanho Yoon[2], Seung-Yeol Lee[3], Hansik Yun[1], Jaebum Cho[1], Kyookeun Lee[1], Hwi Kim[4], Junsuk Rho[2, 5] and Byoungho Lee[1, *]*

[1] School of Electrical and Computer Engineering and Inter-University Semiconductor Research Center, Seoul National University, Gwanak-Gu Gwanakro 1, Seoul 08826, Republic of Korea

[2] Department of Mechanical Engineering, Pohang University of Science and Technology (POSTECH), Pohang 37673, Republic of Korea

[3] School of Electronics Engineering, College of IT Engineering, Kyungpook National University, buk-gu daehakro 80, Daegu 702-701, Republic of Korea

[4] Department of Electronics and Information Engineering, Korea University, 2511 Sejong-ro, Sejong 339-700, Republic of Korea

[5] Department of Chemical Engineering, Pohang University of Science and Technology (POSTECH), Pohang 37673, Republic of Korea

**Abstract:**

Reconstruction of light profiles with amplitude and phase information, called holography, is an attractive optical technique to display three-dimensional images. Due to essential requirements for an ideal hologram, subwavelength control of both amplitude and phase is crucial. Nevertheless, traditional holographic devices have suffered from their limited capabilities of incomplete modulation in both amplitude and phase of visible light. Here, we propose a novel metasurface that is capable of completely controlling ***both amplitude and phase profiles of visible light independently with subwavelength spatial resolution***. The simultaneous, continuous, and broadband control of amplitude and phase is achieved by using X-shaped meta-atoms based on expanded concept of the Pancharatnam-Berry phase. The first experimental demonstrations of complete complex-amplitude holograms with subwavelength definition are achieved and show excellent performances with remarkable signal-to-noise ratio compared to traditional phase-only holograms. Extraordinary control capability with versatile advantages of our metasurface paves a way to an ideal holography, which is expected to be a significant advance in the field of optical holography and metasurfaces.

---

*corresponding author email: byoungho@snu.ac.kr

Holography is an optical technique that reconstructs the wavefront of electromagnetic waves with both amplitude and phase information to display three-dimensional (3D) images [1]. Holography has been an attractive issue since it was used to realize ultimate 3D displays or optical data storage [2–4]. However, conventional digital holography techniques, which are based on typical optoelectronic devices such as spatial light modulators and digital micromirror devices, have been suffering from several issues due to their large pixel size compared to operating wavelengths: low resolution, narrow viewing angle, and severe noises generated from undesirable diffraction orders and twin images. More importantly, conventional holography techniques are usually based on the amplitude- or phase-only modulation scheme with incomplete approximations of object images. In principle, a complex-amplitude modulation, which means a modulation of light capable of both amplitude and phase information independently, is required to perfectly reconstruct the profile of light. For conventional devices, however, operating mechanisms of both amplitude- and phase- modulation are fundamentally interfering with each other, which causes challenges for independent and full control of them. Even if there were a few reports on complex-amplitude modulation schemes applying the conventional optoelectronic devices, they have still suffered from issues including twin image generation, huge optical system size, and limited modulation ranges [5–9]. Hence, in spite of its significance, implementation of high quality complex-amplitude modulation scheme has not been demonstrated yet.

Metasurfaces, which are planar optical elements with composition of artificially fabricated photonic atoms, have attracted extensive interest owing to their numerous functionalities and potentials to modify electromagnetic characteristics [10–14]. The applied areas of metasurfaces have been rapidly expanding with noticeable examples including negative index materials and unusual nonlinear optical materials [15–19]. Recently, metasurfaces are expected to pave the path for high quality spatial light modulators that can overcome the limitations of conventional optical components. A great deal of metasurfaces have been investigated as holographic devices with control of phase [20–30], polarization [31], and both components of light [32–35]. Furthermore, some studies attempted to develop a simultaneous control of amplitude and phase of light; Plasmonic metasurfaces with V-shaped meta-atoms were proposed, but with limited modulation ranges, sophisticated fabrication requirements as well as low efficiency [36]. Metasurfaces with C-shaped meta-atoms were also demonstrated with terahertz waves [37], but fabrication feasibility with their convoluted modulation schemes makes it difficult to implement scalable design of them for visible light. Other types of metasurfaces were also suggested, but within theoretical considerations with low practicality or restricted capabilities [38–40]. As a result, none of the metasurfaces achieves complete control of complex-amplitude at visible range with subwavelength resolution and high efficiency.

In this article, we propose and experimentally demonstrate an advanced holographic metasurface that enables full complex-amplitude modulation of visible light with subwavelength spatial resolution, which can achieve the most complete holograms reaching to ideal holography. As a building block, an X-shaped meta-atom is introduced based on the expanded concept of the Pancharatnam-Berry phase. The X-shaped meta-atom, which is made of poly-crystalline silicon (Poly-Si), enables that full ranges of both amplitude and phase can be mapped and tailored by tuning the orientations of its two arms. Furthermore, the metasurface provides broadband and chiral operations due to the great feature of the Pancharatnam-Berry phase [17, 20, 22, 25, 27, 29, 31, 42–49]. Due to simple and intuitive design strategy, proposed metasurface is rigorous to fabrication errors and shows high efficiency at mid-visible region (40 % at 532 nm). To the best of our knowledge, this is the first time ever to experimentally realize the completely complex-amplitude modulated holograms for visible light, which is expected to be a significant advance in optical holographic technology and metamaterials.

**Theory of metasurface with X-shaped meta-atoms**

A schematic depicting the mechanism of the X-shaped meta-atoms is represented in Fig.1a. First of all, let us start by explaining the key principle of the proposed meta-atom, the Pancharatnam-Berry phase. The Pancharatnam-Berry phase, or geometric phase, is one of the most useful phenomenon to describe the phase profile of scattered light by the spin-rotation coupling. The geometric-phase metasurfaces commonly consist of a unit-cell of a rectangular nanorod, which are composed of plasmonic or dielectric materials. As shown in the left side of Fig. 1a, electric dipole moments are induced parallel to the major axis of the nanorod when the circularly polarized light ($\sigma$) is normally incident to the nanorod. Here, the parameter $\sigma$ is selected as +1 or -1 for right or left circular polarization, respectively. The orientation angle ($\theta_1$ with respect to *x*-axis) of the nanorod leads to time delay of the dipole excitation, which makes relative phase delay among the nanorods according to their own orientation angles. As a result, the scattered light with opposite handedness (-$\sigma$) experiences relative phase shift ($2\sigma\theta_1$), which is only proportional to the orientation angle. The geometric phase is a purely geometrical effect, which can lead to broadband characteristics whose tendency of the phase shift is independent of the wavelength of light.

In this context, the concept of geometric phase can be expanded. If two nanorods with different orientation angles ($\theta_1$ and $\theta_2$) are overlapped, they will construct the X-shaped structure whose arms direct to the angle of $\theta_1$ and $\theta_2$. Assuming that the X-shaped structure can be modeled by two independent electric dipoles, the behavior of the X-shaped structure is allowed to be analyzed by the superposition of the geometric phase. The phase components of cross-polarized waves scattered by these electric dipoles are just proportional to their orientation angles, but with the same amplitude components. Hence, the complex-amplitude of the cross-polarized wave ($E_{cross}$) radiated from the X-shaped structure can be simply expressed as,

$$E_{cross} \propto \frac{1}{2}(e^{j2\sigma\theta_2} + e^{j2\sigma\theta_1}) = \frac{1}{2}(e^{j\sigma(\theta_2-\theta_1)} + e^{-j\sigma(\theta_2-\theta_1)})e^{j\sigma(\theta_2+\theta_1)} = \cos(\theta_2-\theta_1)e^{j\sigma(\theta_2+\theta_1)} \qquad (1)$$

According to Eq. 1, complex-amplitude $E_{cross}$ can be separated in the amplitude and phase component. Amplitude $A$ and phase $\phi$ of total cross-polarized wave can be expressed as $\cos(\theta_1-\theta_1)$ and $\sigma(\theta_1+\theta_2)$, respectively. That is, the amplitude of scattered light can be determined by a difference between two orientation angles while the phase of scattered light can be determined by a summation of two orientation angles, so we confirm that the complex-amplitude modulation can be realized from the double electric dipoles. In the case of actual X-shaped structure, however, higher order modes rather than the electric dipole mode, such as magnetic dipoles and electric quadrupoles, can be excited when a circularly polarized light illuminates on the backside of the structure. However, we found that these higher order modes can be suppressed and only double electric dipoles are dominantly generated near the specific resonance. This resonance can occur with accurately designed geometric parameters such as thickness, length, and width of the X-shaped structure. Therefore, the proposed X-shaped structure can modulate incoming light with the full coverage of complex-amplitude domain according to Eq. 1.

Based on the theoretical analysis, the X-shaped meta-atoms are implemented by using the dielectric materials. A schematic diagram of a proposed unit cell is shown in Fig. 1b. The meta-atom consists of Poly-Si on a glass substrate with the thickness of $t$. The metasurface is then composed of periodically arranged square lattices of X-shaped meta-atoms with the period $P$ for both *x*- and *y*- directions. The proposed structure is designed to operate on the visible light with the wavelength of 532 nm, so the period ($P$ = 350 nm) is set to be shorter than the wavelength of light. Since the period is even shorter than the wavelength in the glass substrate with refractive index of 1.45, there are no diffraction orders in both transmission and reflection. The thickness of Poly-Si ($t$ = 128 nm) is chosen for composing a Fabry-Perot resonator with a low quality-factor, which not only enhances the modulation efficiency but also allows broadband operations. Figure 1c shows the top view of the unit cell. Each pixel has its own orientation angles ($\theta_1$ and $\theta_2$) about the *x*-axis. The angular disparity between the orientation angles is defined as *α*. Considering both fabrication feasibility and modulation efficiency, other parameters such as the length of the nanorod ($L$ = 280 nm) and the width of the nanorod ($w$ = 65 nm) are also carefully designed. Although only a transmission-type metasurface is discussed here, we have confirmed that a reflection-type metasurface can also be implemented in the same manner (see Supplementary Section 1).

**Verification of the analytical approach with full-wave simulations**

The capability of the X-shaped meta-atom is verified by a commercial tool (COMSOL) based on the finite element method (FEM). Detailed information about simulations is explained in Methods. Electric field maps of the X-shaped meta-atom with normal incidence of circular polarization are firstly calculated, which are shown in Fig. 1d. Two cases of X-shaped structures with α=60° and 90° are considered as examples here (for other cases, see Fig. S4). In the figure, colours represent the real part of cross-polarized electric fields ($E_{-\sigma}$) while the black arrows represent the magnitude and directions of electric polarization vectors. It is noticeable that we allow the values of *α* to be from 60° to 90° because a meta-atom having *α* smaller than 60° is not appropriate to apply the superposition mechanism of X-shaped structure due to severe overlap of two nanorods, which is not suitable to our theoretical approach. As shown in Fig. 1d, it can be confirmed that the electric dipoles are dominantly induced along the arms of the X-shaped meta-atom in the defined range, which means that the assumption of our theoretical approach is practically valid.

Figure 2 shows cross-polarized transmission coefficient ($t_{cross}$) as a function of the orientation angle ($\theta_1$) and the angular disparity ($\alpha$) calculated from both theoretical calculations and FEM simulations. Here, $t_{cross}$ is defined as the ratio of complex-amplitude of the cross-polarized transmission ($E_{cross}$) to incident light ($E_{in}$) as depicted in Fig. 2a. Figures 2b–2d present the results of theoretical calculations based on Eq. 1 that models the X-shaped meta-atom as double electric dipoles whose directions are parallel to the major axis of two arms. Corresponding FEM simulation results for the actual X-shaped meta-atom are presented in Figures 2e–2g. In order to compare the results in detail, they are represented separately in each component such as the amplitude (Figs. 2b and 2e) and phase (Figs. 2c and 2f). Figures 2d and 2g describe all of the $t_{cross}$ in the complex domain whose $x$- and $y$-axis mean the real and imaginary part of the complex-amplitude, respectively. As expected, both results of modeling and simulations show strong agreement. The angular disparity is related with the amplitude, whereas the orientation angle is related with the phase relying on the Eq. 1. In addition, the simulation indicates that the maximum efficiency, which is defined as the efficiency for the maximized amplitude, reaches 49 % at the wavelength of 532 nm. As a consequence, we can confirm that our theoretical approach is fairly accurate, and both amplitude and phase of transmitted and cross-polarized light can be fully described by the orientation angles of two arms of X-shaped meta-atom with high efficiency.

**Experimental demonstrations of X-shaped metasurfaces**

To experimentally demonstrate extraordinary capability and versatility of the proposed metasurface, metasurfaces composed of the X-shaped meta-atoms are fabricated according to the geometric parameters by standard electron beam lithography process (see details in Methods). In this section, we characterized two different categories of the metasurfaces with a target wavelength of 532 nm. The first category is a set of metasurfaces with a periodic array of identical X-shaped meta-atoms to measure transmission characteristics of the unit cell. The other category is demonstrated for validating the performance of full complex-amplitude holographic metasurfaces using X-shaped meta-atoms compared to a phase-only holographic metasurface.

Four devices belonging to the first category consist of identical X-shaped meta-atoms with four types of angular disparities $α = 60°$, $70°$, $80°$, and $90°$, respectively. Figures 3a–3d represent their field-enhanced scanning electron microscope (FE-SEM) images. According to the images, it is possible to confirm the feasibility of the proposed structures at the nanoscale. The samples were then illuminated from the bottom by a laser with the free-space wavelength ($λ_d$) of 532 nm. On the transmitted side, a cross-polarized circular analyzer comprising a quarter waveplate and a linear polarizer is used for filtering and measuring the cross-polarized component of transmitted light. By using the oppositely directional circular analyzer comprising a quarter waveplate and a linear polarizer, cross-polarized component of transmitted light can be measured. The results are shown in Fig. 3e with simulation results. The intensity values in Fig. 3e are normalized by the case of $α = 60°$. According to the graph, both experiments and simulations indicate that the intensity of $t_{cross}$ sinusoidally decreases when α increases from 60° to 90°, and finally it goes to zero at $α=90°$. For the maximum case of the experimental results, we measured a maximum efficiency of 40% in cross-polarized transmission. The measured efficiency is a bit smaller than its corresponding simulated values (49 %) because of slight differences between the geometry of the designed and fabricated structures.

The second category of our metasurfaces is prepared for validating functionalities of full complex-amplitude metasurface holograms. As mentioned earlier, extraordinary capability of X-shaped meta-atoms can satisfy the requirement of ideal complex-amplitude holograms. We designed the computer-generated holograms (CGHs) for right circularly polarized light with normal incidence. As shown in Fig. 4a, the CGHs are designed to generate letters "SNU" in 3D space. The alphabets "S", "U", and "N" are displayed on different image planes in the Fresnel region, which are at $z$ = 0 μm, 80 μm, and 150 μm, respectively. It is noticeable that one of the image planes is located at $z$ = 0 μm, which is directly on the metasurface plane, to show the unique performance of complex-amplitude holograms. For conventional phase-only holograms, only phase information of object images remains, whereas the amplitude information of them is set to identical. Therefore, it is impossible to describe the images directly on the metasurface plane. On the other hand, complex-amplitude holograms have both amplitude and phase information, which is capable of describing the reconstructed images on arbitrary planes including the surface of metasurfaces. Therefore, the proposed example can definitely prove whether given metasurfaces modulate phase, amplitude, or both. Calculated CGHs with a sample size of 210 μm × 210 μm and a pixel size of 350 nm × 350 nm are shown in the upper side of Fig. 4d. Any approximations or algorithms that should be used in conventional phase-only holograms, such as random phase injection, are not employed to obtain the expected amplitude and phase profiles, and the back-propagation of the desired image profiles is taken into account according to the Fresnel diffraction theory. The optical microscopic image and FE-SEM images of the fabricated devices possessing the designed CGHs are shown in Fig. 4b and 4c, respectively.

Figure 4d shows the captured holographic images at each *z*-plane in both simulations and experiments. Optical microscopic setup with the cross-polarized analyzer was used to measure the holographic images. Details about the experiments are explained in Methods, and the illustration of the setup is shown in Fig. S1. It is noteworthy that the holographic images of both simulations and experiments have almost the identical profiles without any significant noises, demonstrating its great capability of holographic images in 3D space. Moreover, the letter "S" reconstructed on the image plane very close to the metasurface is purely described, which indicates that the amplitude profiles are correctly implemented as well as the phase profiles. A signal-to-noise ratio (SNR), which is defined as the ratio of the maximum intensity in the holographic image to the standard deviation of the background noise [36], was used to evaluate the image quality. For the experimentally reconstructed image in Fig. 4d, the SNR is 211.3 where the background area was set to the size of 32 μm × 32 μm. This is a remarkable record where the SNRs for previous works have only the values around 50 even though they used approximation algorithms that improve the image quality, but with sacrificing the original wavefronts [36, 44].

For comparison, phase-only metasurface with uniform amplitude profiles is also designed where the other conditions are the same as those of the sample in Fig. 4d. Due to the deficient expressiveness of phase-only holograms, multiplication of randomly-valued phase profiles should be employed to improve the image quality of the phase-only holograms (see details in Methods). Calculated phase profile of the phase-only CGH is shown in the upper side of Fig. 4e. Both calculated and measured holographic images in Fig. 4e show that the letter "S" cannot be formed at the metasurface plane as well as the other letters "N" and "U" are displayed, due to severe noises around images. Comparing the holographic images of Figs. 4d and 4e, we conclude that the complex-amplitude hologram can overcome the speckle noise problem of typical phase-only holograms, significantly expand the range of reconstructed image plane, and simplify the complicated calculation processes of the phase-only CGHs often involving iterative algorithms.

One of the attractive properties of the proposed metasurface is a broadband characteristic. As proved in theoretical approach, the nature of the X-shaped meta-atoms originates from the geometric phase, which only depends on the orientation of the structure. In addition, the broadband characteristic of the proposed metasurface is more improved by using all-dielectric materials, which have less sensitive resonance properties compared to plasmonic materials with sharp plasmonic resonances [41]. Therefore, it is expected that phase and amplitude of cross-polarized transmission are relatively independent of the wavelength of the incident light. We performed the simulations to analyze the structures at different wavelengths (See Fig. S5). Consequently, the phase is absolutely independent on the wavelength, and the amplitude also relies on the theory, but with a few fluctuations relying on the orientation of the structure for wavelengths longer than $\lambda_d$. These fluctuations are due to the interactions among adjacent unit cells where the interactions are hard to be ignored for different wavelengths, which can be sufficiently corrected by scale optimizations of the geometry for desired wavelength regions. By the way, the main tendency of the complex-amplitude modulation is conserved for the different wavelengths. The metasurface used for demonstrating a hologram shown in Fig. 4d was reused for experiments with different wavelengths of 660 nm and 473 nm, and the images at the preferred *z*-planes were captured as shown in Fig. 5. As expected, the holographic images are well reconstructed at each *z*-plane, while just *z*-positions of image planes are changed with respect to the operating wavelengths. The positions of the images can be theoretically calculated according to the principle of geometrical optics, and measured positions of image planes are well relying on their corresponding calculations in all wavelengths.

**Conclusions**

In summary, we have proposed a new type of metasurface composed of X-shaped meta-atoms to realize full complex-amplitude modulation at broadband visible wavelength region. Theoretically, the concept of geometric phase is expanded and applied to materialize an X-shaped meta-atom. Theoretical discussions and numerical simulations of the X-shaped meta-atom have verified that both phase and amplitude of light can be spatially controlled in the range of 0–1 for amplitude and 0–2π for phase within a subwavelength resolution (350 nm × 350 nm). Experiments with regularly patterned metasurfaces show a feasibility and a precise modulation capability of the X-shaped meta-atom. The maximum efficiency of 49 % in simulation and 40 % in experiment for cross-polarized transmission is achieved at the wavelength of 532 nm. Our selection for the material, Poly-Si, makes benefits for high efficiency as well as standard fabrication processes with low cost. Applying the metasurface to transmission-type and on-axis holograms, a complete hologram is realized experimentally with arbitrary complex-amplitude profiles at broadband visible wavelengths. The extraordinary capability of the complex-amplitude hologram is also discussed compared to phase-only hologram. According to the results, it is confirmed that the complex-amplitude hologram can provide excellent 3D images and absolutely eliminate the speckle noise problems of classical hologram where the experimental SNR is valued at 211.3.

A great advantage of the X-shaped meta-atom is its applicability for extensive fields of metasurface platforms based on the geometric phase. Broadband and chirality of the geometric phase have been applied to numerous functional metasurfaces. Polarization-multiplexed holographic imaging [42–44], chiral and multispectral imaging [45], and multi-functional flat optic devices [31, 46–49] are achieved by using the nature of the geometric phase. Because the X-shaped meta-atom originates from the geometric phase, it is worth mentioning that proposed X-shaped meta-atom can also be applied to all of the previous approaches using the nature of geometric phase. In addition, it can improve their performances and have a potential to add a novel degree of freedoms on those applications. Although the visible light is mainly discussed in this article due to its significance on a holographic imaging, the operation range of the metasurface is scalable to other wavelengths such as near-, mid-infrared, and terahertz spectra.

We conclude that the modulation capability of the proposed metasurface as well as its aforementioned advantages almost reach to ideal 3D holograms. We believe that our technology paves the way to advanced holographic display systems, which have been expected to be a next generation display [2–4]. Moreover, our metasurface platform can be applied to not only metasurface platforms but also wide spread optical applications. Our extraordinary degree of freedom for controlling electromagnetic waves will expand the superior limits of various optical applications. Besides the potential applications including arbitrary beam shaping, 3D biological imaging, optical computing, integrated fibre-optics, and nanolasers, "Optics-on-a-chip" has been started with this research. Ultimately, a perfect control of light, which means full and simultaneous control of amplitude, phase and polarization of light with subwavelength resolution, is no longer a dream.

**Method**

**Numerical simulation**. The results in Figs. 1d and 2 were performed by commercial software (COMSOL Multiphysics 5.0) based on the finite element method (FEM). In the simulation, the periodic boundary conditions were used with the period of 350 nm. The material indices constructing the structures are based on the reference [50]. In Fig. 1d, electric fields with polarization vectors are calculated on the *xy*-plane at the upper surface of the X-shaped meta-atom. For Fig. 2, The transmission coefficients of the proper polarizations are computed for all combined values of the orientation angles $\theta$ and the angular disparity α in the range from 0° to 180° and from 60° to 90°, respectively. The period of the unit cell is set to 350 nm.

The simulation results in Figs. 4d and 4e for holographic images were calculated by using Fresnel diffraction theory. For these simulations, the input light is assumed as an ideal plane wave with normal incidence. Desired phase and amplitude profiles represented in the upper side of Fig. 4d are then multiplied to the plane wave, and the *z*-directional propagations of them are calculated for proper image planes by using the angular spectrum methods [1].

**Design of Holograms**. The angular spectrum method (ASM) was used to calculate computer generated holograms (CGHs). The ASM is a method based on the diffraction optics to calculate diffracted light fields by decomposing light field into plane waves. In the process of ASM, there is no paraxial approximation, so it is appropriate to calculate holograms with wide bandwidth. Target images consist of three letters of "SNU", of which each alphabet is at different depth planes. To calculate, each letter image at the different depth is back-propagated to the metasurface plane by ASM. By integrating all back-propagated profiles on the metasurface plane, finally, the CGH can be well calculated because ASM is well defined between parallel planes.

Here, continuous and full ranges of amplitude and phase were used to construct the CGH without any discrete levels, which makes more clear holographic images. Due to the advantage of full complex-amplitude modulation capability, CGH is calculated without any support of optimization algorithms such as the Gerchberg-Saxton (GS) algorithm.

The phase-only CGH in the upper side of Fig. 4e was also calculated for the comparison with the full complex CGH. The phase-only CGH was calculated from the complex CGH by flattening its amplitude profiles. For the fair comparison, additional optimization algorithms such as the GS algorithm were not applied to the phase-only CGH. The GS algorithm can improve an accuracy of amplitude profiles on the image plane. However, this algorithm causes that the phase profile on the image plane, which should be a degree of freedom, cannot be controlled as desired. It is worth noting that this is the critical limitation of GS algorithm for practical three-dimensional holograms. A target wavelength is set to 532 nm, and the sampling period of the CGH is set to 350 nm. All the fabricated devices in this research have identical pixel numbers (600 × 600 pixels).

**Device fabrication**. Standard electron-beam lithography processes with lift-off and etching processes were used to fabricate the metasurfaces. First, an intrinsic poly-crystalline silicon (Poly-Si) film with 128 nm thickness was deposited on a fused silica wafer using low-pressure chemical vapor deposition (Eugene Technology BJM-100) at the temperature of 700℃. Then, an electron-beam lithography (Elionix ELS-7800) and a standard lift-off process were used to create the designed X-shaped patterns with chromium (Cr) hard-mask layer. Inductive coupled plasma based reactive ion etching (ULVAC NE-7800) was employed to etch the Poly-Si layer along the patterned Cr, and Cr masks were then removed by using chromium etchant (KMG CR-7) after all of the processes.

**Measurement**. A schematic of the optical set-up is depicted in Fig. S1. A high-power optically pumped semiconductor laser (OPSL; Coherent, Verdi G2 SLM) with a wavelength of 532 nm illuminates the back side of the metasurfaces after passing through a spatial filter, a half-wave plate (HWP), a quarter-wave plate (QWP), and an iris. To generate well-collimated beams, a beam spatial filter was used to eliminate the high orders of the Gaussian beam. The HWP and QWP were employed to set the polarization of incident beam into the right circular polarization state, and the iris was for restricting multiple reflections among the optic elements. A visible polarimeter (Thorlabs, PAN5710VIS) was used to precisely generate the proper polarization state. The metasurfaces samples and the lens are mounted on XYZ stages to carefully steer the relative positions on the beam path. In the measurements, the beam radius was set to be about 500 μm, which is fully larger than the size of metasurfaces (~210 μm). The optical microscopy, which consists of the objective lens and the tube lens, was applied to measure the images at proper image planes, and the visible charge-coupled device (CCD) camera captured images.

To measure the results in Fig. 5, the laser in the schematic of Fig. S1 was exchanged to other optical lasers of which wavelengths are 660 nm (Cobolt, Flamenco 300) and 473 nm (Spectra-Physics, Excelsior 473) for red and blue colours, respectively.

To obtain the maximum efficiencies of the metasurfaces, the same optical setup in Fig. S1 is used except the CCD camera, which is exchanged to the set of pinhole and a power meter device. The optical power passed through the metasurface and the pinhole was measured using the power meter and divided by the power of the incident beam.

# FIGURES

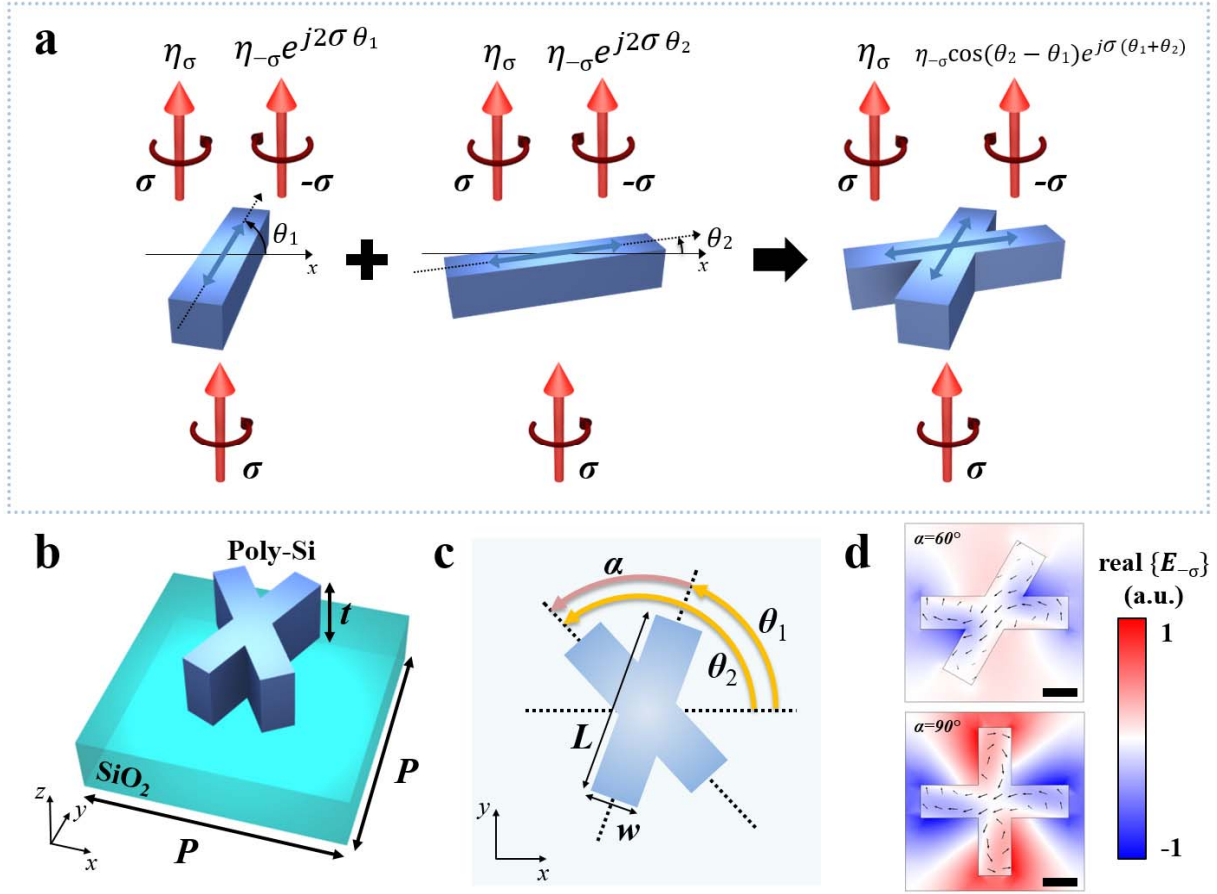

**Figure 1. Schematic illustrations of an X-shaped meta-atom** (a) Illustration describing the operating mechanism of the structures. Rotation of the single nanorod with orientation angle $\theta_1$ (or $\theta_2$) leads to the phase variance $2\sigma\theta_1$ (or $2\sigma\theta_2$) in cross-polarized component of transmitted light according to the Pancharatnam-Berry phase. The superposition of the two nanorods with different orientation angles $\theta_1$ and $\theta_2$ implements the X-shaped structure, which has amplitude $\cos(\theta_2-\theta_1)$ and phase $\sigma(\theta_1+\theta_1)$ in cross-polarized component of transmitted light. $\eta_\sigma$ and $\eta_{-\sigma}$ are coupling coefficients of co- and cross- polarized light, respectively. (b) Schematic of a unit cell of the proposed metasurface. An X-shaped meta-atom consists of Poly-Si on a glass substrate with thickness $t$ and the unit cell period $P$. (c) Top view of the unit cell showing longer length $L$, width $w$, orientation angles of its arms $\theta_1$ and $\theta_2$, and their disparity $\alpha$. For the metasurfaces designed at operating wavelength $\lambda_d$=532 nm, X-shaped meta-atoms have $L$=265 nm, $w$=65 nm, $t$=128 nm, and $P$ = 350 nm. (d) Simulation results that show the distribution of electric fields and polarization vectors. Colours represent the real part of cross-polarized electric fields ($E_{-\sigma}$), and black arrows in the figures represent the magnitudes and directions of the polarization vectors at each position. The upper figure has the angular disparity $\alpha=60°$ while lower figure has $\alpha=90°$. Other geometric parameters of both structures are identical. Scale bar, 50 nm.

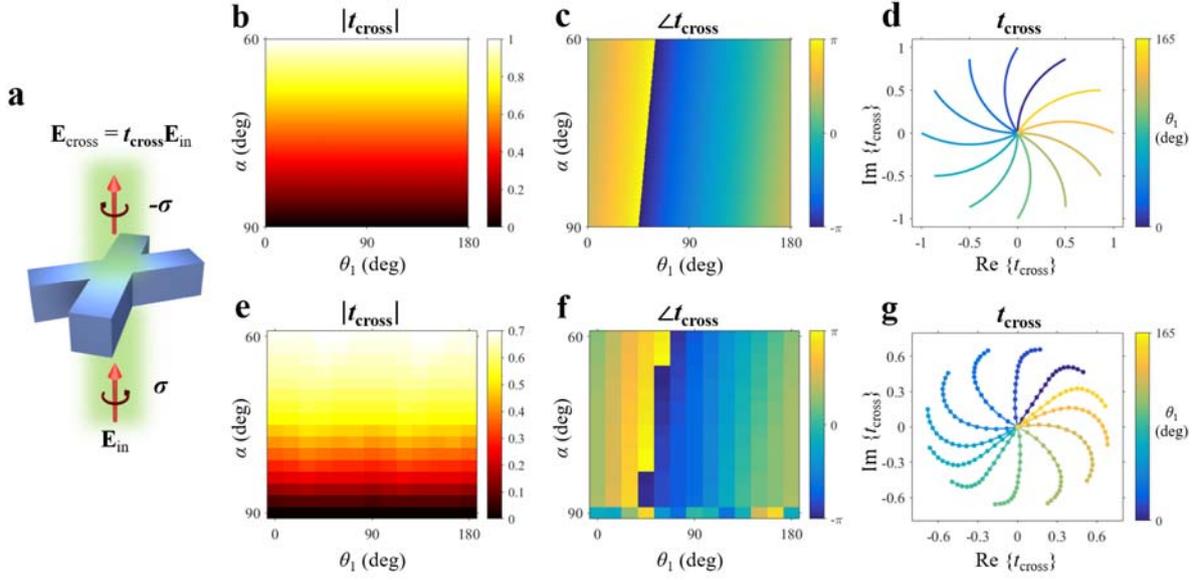

**Figure 2. Theoretical and Numerical analysis on cross-polarized transmissions of an X-shaped meta-atom.** (a) Schematic illustration of a unit cell of an X-shaped meta-atom. Cross-polarized transmission coefficient $t_{cross}$ is defined as the ratio between the complex amplitude of cross-polarized component of transmitted light and that of incident light. Results of (b–d) analytical calculations and (e–g) FEM simulations for $t_{cross}$. (b, e) Variation in the amplitude component of $t_{cross}$ for different orientation angles $\theta_1$ and disparities $\alpha$. (c, f) Variation in the phase component of $t_{cross}$ for different orientation angles $\theta_1$ and disparities $\alpha$. (d, g) Accessible range of $t_{cross}$ plotted in the complex domain. The different colours of the lines represent the cases of different orientation angles, and the colour bar is represented on the right side of the graphs.

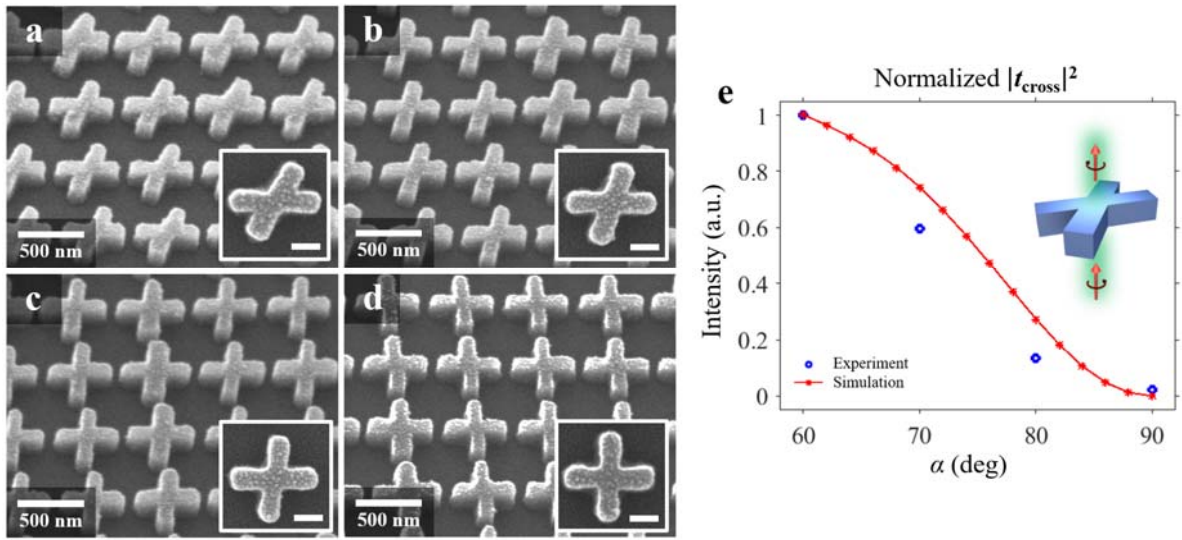

**Figure 3. Experimentally measured cross-polarized transmissions ($t_{cross}$) of the X-shaped meta-atoms in several angular disparities.** (a–d) FE-SEM images of the metasurfaces composed of regularly patterned X-shaped unit-cells with angular disparities $\alpha$=60°, 70°, 80°, and 90° for (a), (b), (c), and (d), respectively. Scale bar of inset, 50 nm. (e) Results of both simulations and experiments for the cross-polarized transmittance as a function of the angular disparity $\alpha$. Blue circles indicate the experimental results while the red crossed points and curve are for the FEM simulation results. The results are normalized to the transmittance for the case of $\alpha = 60°$ which is designed to have the maximum transmittance.

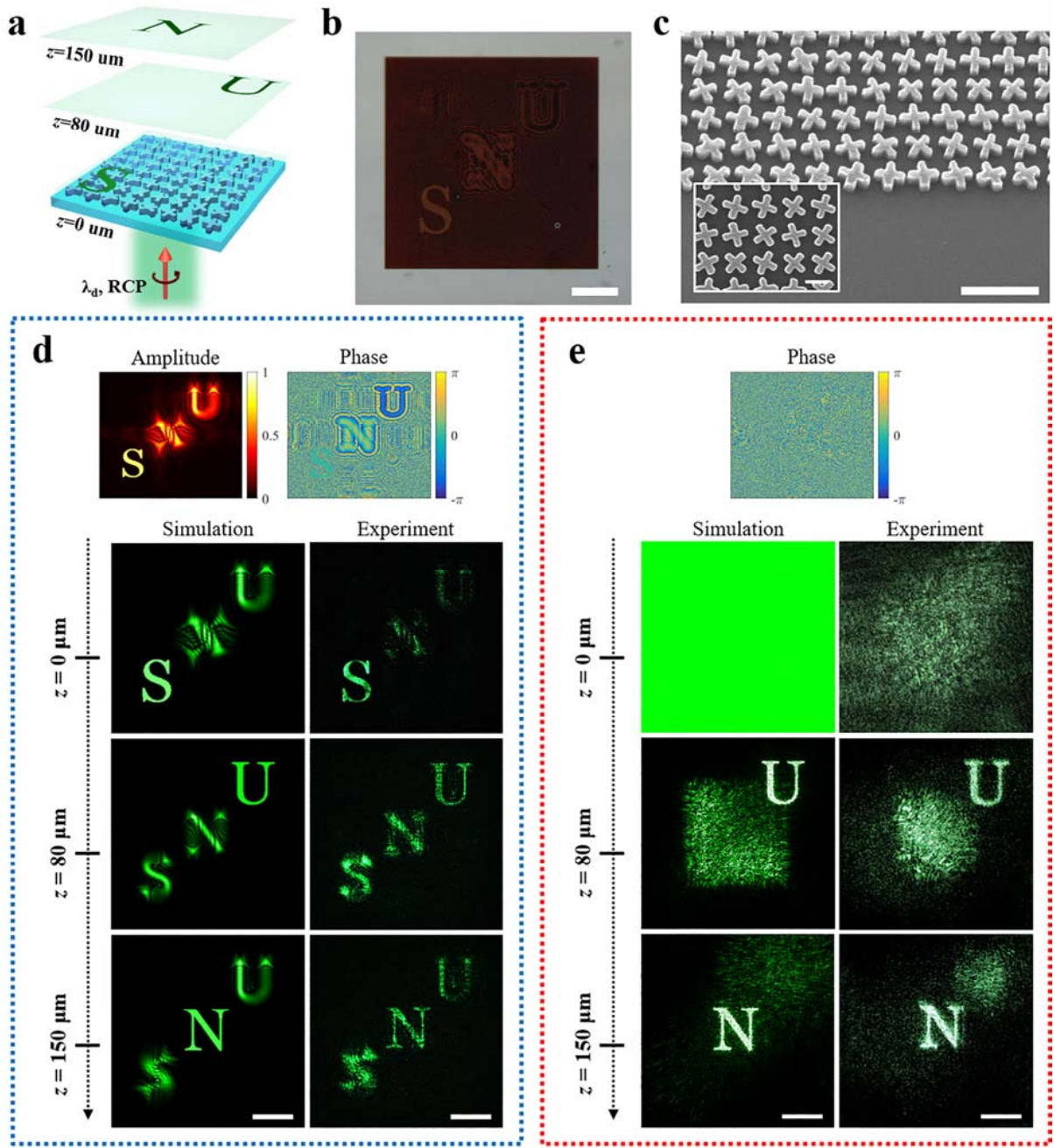

**Figure 4. Full complex-amplitude and phase-only holograms reconstructed by X-shaped metasurfaces.** (a) Schematic diagram for the designed hologram. The right circularly polarized light with wavelength $\lambda_d$=532 nm was used to reconstruct the hologram. Three letters "SNU" is reconstructed on the proper image planes that are on the metasurface plane for "S", *xy*-plane on the $z = 80$ μm for "U", and *xy*-plane on the $z = 150$ μm for "N", respectively. (b) Optical microscopic image of the fabricated device on the fused silica wafer. Scale bar, 50 μm. (c) FE-

SEM images of the fabricated metasurfaces for the tilted view and top view (inset). Scale bars are 500 nm and 300 nm for the main and inset images, respectively. (d) Designed CGH and captured holographic images of complex-amplitude hologram. Amplitude and phase profiles of CGHs with 600 × 600 number of pixels are represented on the upper side. Calculated images and their corresponding measured intensity distributions at proper image planes are shown in lower side. Continuous change of the holographic images with respect to $z$ is shown in Movie S1 with explanation in Fig. S6. Scale bar, 50 μm (e) Designed phase-only CGH and captured holographic images of phase hologram. Calculated and measured intensity distributions are shown in lower side while the phase profile is shown in the upper side. Scale bar, 50 μm. Experimental images are captured by a colour CCD camera.

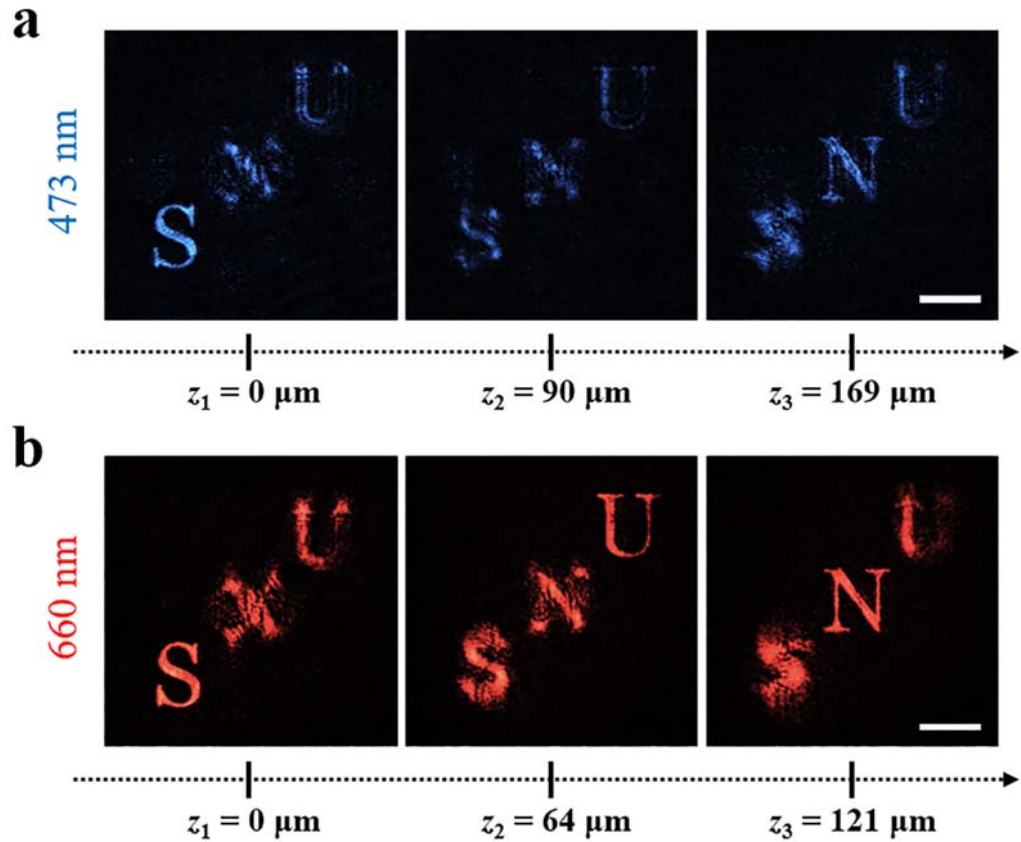

**Figure 5. Experimentally captured holographic images for verifying broadband characteristic of the X-shaped metasurface.** Measured images at proper image planes for the wavelength of (a) 473 nm and (b) 660 nm. Positions of image planes are shown in the figures for each case. Their corresponding values of theoretical calculations at 473 nm are $z_1 = 0$ μm, $z_2 = 90$ μm, and $z_3 = 169$ μm, while the values at 660 nm are $z_1 = 0$ μm, $z_2 = 64$ μm, and $z_3 = 121$ μm. Continuous change of the holographic images with respect to $z$ is shown in Movie S2 and S3 with explanation in Fig. S6 for the wavelength of 473 nm and 660 nm, respectively. Scale bar, 50 μm.

AUTHOR INFORMATION

**Corresponding Author**

*E-mail: byoungho@snu.ac.kr

**Notes**

The authors declare no competing financial interest.

**Author Contributions**

G.-Y.L. conceived the ideas and carried out the design and simulations of the metasurfaces. G.-Y.L. and S.-Y.L. analyzed and discussed the theoretical approaches. G.-Y.L., S.-Y.L. and H.Y. conceived and designed the experiments. J.C. and G.-Y.L. designed and calculated the holograms. G.Y. fabricated the devices, and G.Y., J.R. and G.-Y.L. discussed the fabrication processes. G.-Y.L., J.C and H.Y. performed the measurements. G.-Y.L., S.-Y.L., H.Y., J.C., K.L. and H.K. analyzed the data. G.-Y.L., S.-Y.L., H.Y., G.Y., J.C., K.L. and H.K. prepared the manuscript. B.L. initiated the project. B.L., J.R. and H.K. supervised the project. All authors discussed the results and commented on the manuscript.

**Acknowledgement**

This work was supported by the Center for Advanced Meta-Materials (CAMM) funded by the Ministry of Science, ICT and Future Planning as Global Frontier Research Project (CAMM-2014M3A6B3063710, CAMM-2014M3A6B3063708). The authors acknowledge support from the Brain Korea 21 (BK21) project. J. R. acknowledges financial supports from NRF Young Investigator program (NRF-2015R1C1A1A02036464) and POSCO Green Science Program.